\documentstyle[12pt]{article}
\begin{document}

\title{Linear $\Sigma$ model in the Gaussian wave functional
approximation II: Analyticity of the S-matrix and the effective
potential/action}

\author{Issei Nakamura$^{(1)}$ and V. Dmitra\v sinovi\' c$^{(2)}$ \\
(1) Research Center for Nuclear Physics, Osaka University \\
Ibaraki, Osaka 567-0047, Japan; and \\
(2) Vin\v ca Institute of Nuclear Sciences \\
P.O.Box 522, 11001 Beograd, Yugoslavia }

\maketitle                 

\abstract{We show an explicit connection between the solution to
the equations of motion in the Gaussian functional approximation
\cite{issei01} and the minimum of the (Gaussian) effective
potential/action of the linear $\Sigma$ model, as well as with the
N/D method in dispersion theory. The resulting equations contain
analytic functions with branch cuts in the complex mass squared
plane. Therefore the minimum of the effective action may lie in
the complex mass squared plane. Many solutions to these equations
can be found on the second, third, etc. Riemann sheets of the
equation, though their physical interpretation is not clear. Our
results and the established properties of the S-matrix in general,
and of the N/D solutions in particular, guide us to the correct
choice of the Riemann sheet.
We count the number of states and find only one in each
spin-parity and isospin channel with quantum numbers corresponding
to the fields in the Lagrangian, i.e. to Castillejo-Dalitz-Dyson
(CDD) poles. We examine the numerical solutions
in both the strong and weak coupling regimes and calculate the
K\" all\' en-Lehmann spectral densities and then use them for
physical interpretation.}
%

\section{Introduction}
\label{sec:intro}

In the present paper we extend the study of a chirally invariant,
Lorentz invariant, self-consistent mean-field, variational
approximation, that goes by the name of Gaussian wave functional
approximation \cite{bg80,hat92} to the linear sigma model, that
was begun in Ref. \cite{issei01}.
We have shown in Ref. \cite{issei01} how to ensure chiral symmetry
in the Gaussian approximation method, a major improvement over
previous treatments. A number of questions have remained open
after that paper, however. In particular we have not addressed the
connection between the Gaussian approximation to the canonical
equations of motion and the Gaussian effective potential (EP)
method, that is rather popular in finite temperature applications
\cite{amcam96}. There the meson masses are defined in terms of the
curvature (second derivative) of the effective potential evaluated
at the minimum. It used to be believed that this definition
leads to a violation of the Nambu-Goldstone (NG) theorem
for the $\pi$ fields \cite{sat87}, even in the chiral limit.
This misunderstanding was cleared up in Ref. \cite{okop96}; moreover
it was shown there that one ought not to minimize the effective
potential, which is momentum independent; rather one must minimize
the effective action (EA) which leads to momentum-dependent equations.
The latter mass definition
leads to an {\it equation} to be solved for $m_{\sigma}$.
Yet there have been no attempts to solve this equation in the
literature, save for Ref. \cite{issei01}. Thus, most $m_{\sigma}$
values present in the literature are not acceptable. The equation
for $m_{\sigma}$ is a transcedental one, however, with infinitely
many Riemann sheets and an apparently indeterminate number and
properties of solutions.
Furthermore, the Gaussian method involves
certain auxilliary objects, such as the solutions to the gap
equations (that are often interpreted as meson masses) whose
physical role is also unclear.
Similarly, the canonical Gaussian approach
involves two-body scattering Bethe-Salpeter (BS)
equations that do not seem to appear in the EP approach.

In this paper we shall answer the aforementioned questions and
some others not mentioned above: For example we
show that exactly the same equations for the meson mass appear
in the canonical Gaussian approximation and the Gaussian effective
action approach: indeed these equations are the net result of the
coupled Bethe-Salpeter and the gap equations. Of course,
this fact does not make them any easier to solve, but it offers a
useful perspective on the number and nature of the solutions.
Inhomogeneous BS equations are scattering equations, and in this
particular approximation they will be shown to be equivalent to
N/D equations of dispersion theory \cite{gas66,nishi74,frau63,collins68},
that ensure manifest unitarity. Some properties of their solutions,
such as analyticity, and the physical interpretation of the solutions,
follow from unitarity and causality, and have been known since the
early 1960's \cite{gas66,nishi74,frau63,collins68}. Another well
known property of N/D equations is the arbitrary number of their
solutions: this is the Castillejo-Dalitz-Dyson [CDD]
ambiguity \cite{gas66,nishi74,frau63,collins68}. The Gaussian
approximation is more restrictive than the N/D approximation,
however: all properties,
such as the number of CDD poles and values of subtraction
constants are determined by the gap and BS equations
that are a part of the canonical Gaussian approximation.

In Ref. \cite{issei01} we have
numerically solved the BS equation in the scalar channel on the
real $s$ axis and found multiple solutions for certain
parameter values, and no solutions at all for others. Yet, in the
weak-coupling limit the Gaussian solution is unique and smoothly
connected to the perturbative one. We use the K\" all\' en-Lehmann
spectral representation to show that the ``heavy" solutions do not
have particle-like properties. In consequence of this, we show
that there is only one solution in each spin-parity-isospin channel.

Thus we have established an explicit connection between some previously
separate formalisms, such as the effective potential/action (a.k.a.
Cornwall-Jackiw-Tomboulis (CJT)) method \cite{cjt74}, the Hartree
+ Random Phase Approximation (RPA), and the N/D method in
dispersion theory, as well as shed light on the particle content
in this approximation.

This paper falls into five sections. After the Introduction, in
Sect. II we outline the Gaussian
effective potential/action method and prove its equivalence with the
Gaussian approximation to the canonical equations of motion.
In Sect. III we demonstrate the
latter's connection with the N/D equations of S-matrix theory. In
Sect. IV we show and discuss the numerical solutions to the gap and
the Bethe-Salpeter equations, calculate the K\" all\' en-Lehmann
spectral function and analyze the particle content of the
solutions. Finally in Sect. V we summarize and draw conclusions.

\section{The Gaussian effective potential/action}

We shall use the notation and conventions of Ref. \cite{issei01}.
In Sect. II and III we shall work in the chiral limit
$(\varepsilon = 0)$, so as to
avoid unnecessary complications. Extension to the non-chiral case is
straightforward. Of course,  we use the non-chiral equations in
the numerical solutions in Sect. IV.

The effective potential, and the effective action methods in
quantum field theory (QFT) were popularized in the mid-70's. These two
are objects with certain intriguing theoretical properties: the former
is the generating function for the zero-momentum one-particle
irreducible (OPI) graphs, the latter is the generating functional for
arbitrary momentum OPI graphs \cite{hat92,iz80}.
At first only the one-loop perturbative approximation was
calculated in the $\phi^4$ theory. These
two objects need not be perturbative, however, and first attempts
at their non-perturbative evaluation were made slightly later.

The Gaussian effective potential is a natural product of a
variational calculation based on the Gaussian ground state trial
wave functional \cite{bg80,hat92} of a scalar single component
$\phi^4$ quantum field theory (QFT). The O(N) symmetric effective
potential was calculated e.g. in Ref.  \cite{sat87}.
In another original approach Cornwall, Jackiw and Tomboulis [CJT],
Ref. \cite{cjt74} used certain disconnected (``vacuum")
two-particle irreducible diagrams to define and calculate the
ground state (``vacuum") energy, as per Goldstone's (many-body)
theorem \cite{fw71}.
The resulting vacuum  energy defines a (non-perturbative) effective
potential \cite{iz80}, which together with the kinetic energy defines
the effective action.
When one minimizes the CJT vacuum energy using a particular
variational Ansatz, the resulting minimization conditions, or the
gap equations are equivalent to the ones obtained in the
canonical Gaussian variational approximation.
It is less obvious that the two-body (Bethe-Salpeter) equation in the
latter formalism is equivalent to the mass equation in the former, 
in the case of spontaneous symmetry breaking.
We shall explicate here the proof given in Refs. \cite{dms96,okop96}.

Instead of the Cornwall-Jackiw-Tomboulis [CJT] appproach we follow
Stevenson, All\` es and Tarrach's more direct calculation \cite{sat87}
based on the Gaussian vacuum wave functionals. The latter
authors found the ``vacuum'' (ground state) energy density ${\cal E}
(M, \mu; \langle \mbox{\boldmath$\phi$} \rangle)$ given by Eq. (3.2)
in Ref. \cite{issei01}. By definition \cite{hat92,iz80}, the
effective potential $V_{\rm eff}(m_{i}, \langle  \phi_{i} \rangle)$
is
\begin{eqnarray}
V_{\rm G}(m_{i}, \langle  \phi_{i} \rangle) &=&
{\cal E}(m_{i}, \langle  \phi_{i} \rangle) -
{\cal E}(m_{i}, \langle  \phi_{i} \rangle = 0).
\end{eqnarray}
One may identify the $\hbar I_{1}(m_{i})$ term in Eq. (3.2) in
Ref. \cite{issei01} with the familiar
``zero-point" energy density of a free spinless field of mass
$m_{i}$
\footnote{Indeed, many finite-temperature/density studies
\cite{amcam96} have been based on this observation, as the
zero-point energy and the effective potential point of view offer
an ``obvious" extension of the zero-temperature/density formalism.}.
This seems to imply that such one meson states are
present in the Gaussian approximation to this theory and that
$m_{i}$ are their physical masses.
That is not the case, however, as we shew in Ref. \cite{dms96,vd98}
(see, also later):
The $m_{i}$ are merely auxilliary quantities (variational
parameters) that determine the position of certain particle production
thresholds and the the corresponding branch cuts, but there are no
poles in the propagators at those mass values.

On the other hand it has been shown
in two different ways \cite{dms96,okop96,issei01} that there
are three massless (Nambu-Goldstone) and one massive ($\sigma$)
state (with a mass different from any of the $m_{i}$) in the
Gaussian approximation. Thus the physical content of the Gaussian
approximation, though formally well established, remains one of
its intuitively most confusing aspects. By working out the
connection between various formalisms we shall shed more light on
this issue.

\subsection{The Gaussian effective potential}

We shall use the fact that the effective potential is the
generating function of OPI zero-momentum Feynman diagrams.
In other words,
the $n$-th derivative of the effective potential is the OPI Green's
function evaluated at zero external momenta. Thus  the curvature
(the second derivative) of the effective potential with respect to the
corresponding fields evaluated at the minimum yield the inverse of
the two-point Green functions $\Gamma_{ii}^{-1}$ evaluated at
zero momentum squared $p^{2} = 0$,
\begin{eqnarray}
- \Gamma_{ii}^{-1}(p=0) &=&
\left({d^{2} V_{\rm G}(\langle \phi_{j} \rangle)
\over{d \langle  \phi_{i} \rangle^{2}}}\right)_{\rm min}
\label{e:mass}
\end{eqnarray}
which, in turn, are interpreted as squared masses of the corresponding
states
by way of
\begin{eqnarray}
- \Gamma_{ii}^{-1}(p=0) &=& m_{i}^{2},
\label{e:def}
\end{eqnarray}
thus equating the meson masses with
Eqs. (\ref{e:mass}).
It had been tacitly assumed that such a definition of $\sigma, \pi$
masses
\begin{eqnarray}
m_{\sigma}^{2} &=& \left({d^{2} V_{\rm G}(\langle \phi_{j} \rangle)
\over{d \langle  \phi_{0} \rangle^{2}}}\right)_{\rm min}
\nonumber
\\
m_{\pi}^{2} &=& \left({d^{2} V_{\rm G}(\langle \phi_{j} \rangle)
\over{d \langle  \phi_{i} \rangle^{2}}}\right)_{\rm min},~i=1,2,3 ,
\label{e:gapm}
\end{eqnarray}
is equivalent to the ``single-particle", or ``gap equation" masses
$M, \mu$, i.e.
that $m_{\sigma} = M$ and $m_{\pi} = \mu$, but that assumption leads
to an apparent violation of the Nambu-Goldstone [NG] theorem, since
$\mu > 0$, even in the chiral limit ($\varepsilon = 0$).
This conclusion is incorrect, albeit very common in the
literature. Many studies were devoted to attempts at its
rectification: for example it was
shown that $\mu \to 0$ in the $1/N \to 0$ limit, and that was
supposed to restore the NG theorem, see Ref. \cite{sat87}.
A straightforward evaluation of the derivatives in Eq. (\ref{e:gapm})
yields \cite{okop96}, however,
\begin{eqnarray}
m_{\sigma}^{2} &=&
M^{2} \left({1 + 2 \lambda_{0}
\left[3 I_{MM}(0) - I_{\mu \mu}(0) \right]
- 24 \lambda_{0}^{2} I_{MM}(0) I_{\mu \mu}(0)
\over{1 - \lambda_{0}
\left[3 I_{MM}(0) + 5 I_{\mu \mu}(0) \right]
+ 12 \lambda_{0}^{2} I_{MM}(0) I_{\mu \mu}(0)}}\right)
\neq M^{2}
\label{e:okop0} \\
m_{\pi}^{2} &=& 0
\neq \mu^{2} ~,
\label{e:okop1}
\end{eqnarray}
where $I_{\mu \mu}(p^{2}), I_{MM}(p^{2})$ are logarithmically
multivalued functions of $p^2$ defined by
\begin{eqnarray}
I_{mm}(s)
&=& i \int {d^{4} k \over (2 \pi)^{4}}
{1 \over {\left[k^{2} - m^{2} + i \epsilon \right]
\left[(k - p)^{2} - m^{2} + i \epsilon \right]}}~, \
\label{e:I}
\end{eqnarray}
where $m = M$, or $\mu$,
explicitly  evaluated in Eqs. (3.21), (3.22) and (3.23) in Ref.
\cite{issei01}.

The result Eq. (\ref{e:okop1}), of course, restores the NG theorem,
but it does so almost {\it per fiat}: it gives one no insight into the
mechanism that brought it about, e.g. it tells us nothing about the
sum of Feynman diagrams that leads to it. After all,
each effective potential generates a certain class of loop diagrams
(at zero external momenta) that one may wish to identify.
It is very difficult to see which class of diagrams corresponds to
Eqs. (\ref{e:okop0}),(\ref{e:okop1}), and which Schwinger-Dyson [SD]
equations sum up
that class. The first, brief answer to that question was given
in Ref. \cite{vd98}; now we shall elaborate on it.
Once we have identified the Feynman diagrams one immediately sees
that Eq. (\ref{e:okop0}) is not quite right.
That fact was also recognized earlier, Ref. \cite{okop96}, but
without reference to Feynman diagrams, and the correction
was also given there, but only with a formal
mathematical justification. Here we shall give an explicit Feynman
diagrammatic interpretation of that formal definition.

\subsection{The Gaussian effective action}

We shall use the fact that
the effective action is the generating
functional of one-particle irreducible (OPI) Green functions
\cite{hat92}, i.e.,
\begin{eqnarray}
\Gamma_{ij}^{-1}(x,y) &=&
\left({\delta^{2} S[\langle \phi_{k} \rangle]
\over{\delta \langle \phi_{i}(x) \rangle
\delta \langle \phi_{j}(y) \rangle}}\right)_{\rm min} ,
\label{e:prop}
\end{eqnarray}
where $\delta$ is the functional derivative.
Then the correct definition of the particle mass is the position
of the pole in the two-point Green function, i.e.
\begin{eqnarray}
\Gamma_{ii}^{-1}(p^{2} = m_{i}^{2}) &=& 0
\label{e:def'}
\end{eqnarray}
where
\begin{eqnarray}
\Gamma_{ij}^{-1}(p) &=&
\left({\delta^{2} {\tilde S}_{\rm G}[\langle \phi_{k} \rangle ]
\over{\delta \langle \phi_{i}(p) \rangle
\delta \langle \phi_{j}(0) \rangle}}\right)_{\rm min}, \
\label{e:mass'}
\end{eqnarray}
and ${\tilde S}$ is the Fourier transform of
the (Gaussian) effective {\it action},
\begin{eqnarray}
S_{\rm G}[\langle \phi_{i} \rangle] &=&
\int d^{4}x \left(T - V_{\rm G}\right) ,
\end{eqnarray}
$T = {1 \over 2} (\partial _{\mu} \langle \phi_{i} \rangle)^{2}$ is
the kinetic energy density.
Eq. (\ref{e:def'}) shows the distinction
between the effective action and the effective potential methods:
the effective potential is momentum independent and thus cannot
correctly describe the pole in the propagator, except when the pole
happens to be at zero momentum/mass. Thus the mass obtained from the
effective potential method agrees with the one obtained from the
effective action only when the mass vanishes.
Therefore we must solve only Eq. (\ref{e:def'}) for the mass,
as the latter appears on both sides of the equation.
This distinction is insignificant for massless (NG) states and its
importance increases with the mass: once
the mass crosses the lowest (particle pair) production threshold, it
acquires an imaginary part that cannot be neglected. Thus the pion
mass calculated via the effective potential might be OK, because it
lies below all hadronic production thresholds, but the scalar meson
mass is definitely not OK.

Equations resulting from Eq. (\ref{e:mass'}) in the linear
$\sigma$ model have been written down, but not solved
(except in the trivial NG pion case) in Ref. \cite{okop96}:
Eqs. (\ref{e:mass'}) yield (N - 1) = 3 massless states
(pions in the chiral limit) and one massive
state ($\sigma$ meson) whose mass $p^2 = m_{\sigma}^{2}$ is
determined by the roots of the following equation
\begin{eqnarray}
p^{2} &=& M^{2} \left(1 + {3 \lambda_{0} \left[3 I_{MM}(p^{2}) +
I_{\mu \mu}(p^{2}) - 12 \lambda_{0}^{2} I_{MM}(p^{2}) I_{\mu
\mu}(p^{2}) \right] \over{1 - \lambda_{0} \left[3 I_{MM}(p^{2}) +
5 I_{\mu \mu}(p^{2}) \right] + 12 \lambda_{0}^{2} I_{MM}(p^{2})
I_{\mu \mu}(p^{2})}}\right) .
\label{e:mass2}
\end{eqnarray}
After making the replacement $p^{2} = m_{\sigma}^{2}$,
and a slight rearrangement Eq. (\ref{e:mass2}) turns into
\begin{eqnarray}
m_{\sigma}^{2} &=&
M^{2} \left({1 + 2 \lambda_{0} \left[3 I_{MM}(m_{\sigma}^{2}) -
I_{\mu \mu}(m_{\sigma}^{2}) \right]
- 24 \lambda_{0}^{2} I_{MM}(m_{\sigma}^{2}) I_{\mu \mu}(m_{\sigma}^{2})
\over{1 - \lambda_{0}
\left[3 I_{MM}(m_{\sigma}^{2}) + 5 I_{\mu \mu}(m_{\sigma}^{2}) \right]
+ 12 \lambda_{0}^{2}
I_{MM}(m_{\sigma}^{2}) I_{\mu \mu}(m_{\sigma}^{2})}}\right) .
\label{e:okop}
\end{eqnarray}
To show equivalence of these results to those of the canonical
Gaussian approximation, we must first remember that in Ref.
\cite{issei01} we showed
the scalar ($\sigma$) sector coupled
Bethe-Salpeter (BS) equations that sum infinite classes of
connected, though not necessarily OPI
Feynman diagrams. The mass of the scalar meson is
determined by the ``pole condition" in the scalar channel BS
solution, Eq. (3.26) in Ref. \cite{issei01}, that reads
\begin{eqnarray}
(s - M^2){\cal D}(s) &=& 0.
\label{e:mass3}
\end{eqnarray}
where
\begin{eqnarray}
{\cal D}(s) &=&  1 - \lambda_{0} \left[
3 \left(1 + 3 {M^{2} \over s - M^{2}} \right) I_{MM}(s) +
\left(5 + 3 {M^{2} \over s - M^{2}} \right) I_{\mu \mu}(s) \right]
\nonumber \\
&+& 2 \lambda_{0}^{2} I_{MM}(s) I_{\mu \mu}(s)
\left(1 + 3 {M^{2} \over s - M^{2}} \right) ,
\label{e:d}
\end{eqnarray}
is the discriminant of the coupled BS equations
(see Ref. \cite{issei01}). Collecting terms we find
\begin{eqnarray}
s &=& M^{2}
\left({1 + 2 \lambda_{0} \left[3 I_{MM}(s) - I_{\mu \mu}(s) \right]
- 24 \lambda_{0}^{2} I_{MM}(s) I_{\mu \mu}(s)
\over{1 - \lambda_{0} \left[3 I_{MM}(s) + 5 I_{\mu \mu}(s) \right]
+ 12 \lambda_{0}^{2} I_{MM}(s) I_{\mu \mu}(s)}}\right)
\label{e:mass1}
\end{eqnarray}
Upon replacing $s = m_{\sigma}^{2}$, this equation becomes
identical to Eq. (\ref{e:okop}) for the $\sigma$ mass. In other
words, the results of the Gaussian effective action approach are
exactly identical to those of the Gaussian BS equation (or, in the
many-body theory language, to the mean-field theory + RPA), proving
which was one of our goals.

As stated above, in the
Gaussian approximation, it is unclear which (classes of) OPI diagrams
are generated by the effective action. We have readily calculated
the analytic form of the effective potential curvature at the
minimum, Eq. (\ref{e:okop}), but it would be a major challenge to
identify the corresponding class of OPI diagrams without the
benefit of the above insights obtained from the Gaussian
Bethe-Salpeter equation in Ref. \cite{issei01}.
Furthermore, the factor $(s - M^2)$ in the equation (\ref{e:mass3})
explicitly shows that the $M$-particle pole has been ``amputated"
from the amplitude,
i.e., that the summed diagrams are one-$M$-particle irreducible.
Some confusion arose due to the fact that $\Gamma_{ij}(p^{2} = s)$
is a
two-point OPI Green function, whereas the BS equation defines a
connected four-point Green function. There is no  contradiction,
however, as one can see after ``amputating" the external ``legs"
of the BS amplitude: the result is just an $s$-channel propagator,
i.e., a two-point Green function.
We have thus given an explicit proof of a formal property
of the effective action, but this does not begin to tell us what
branch of Eq. (\ref{e:okop}) to solve.

As the solutions to Eq. (\ref{e:okop}) are expected to lie in the
complex $p^{2}$ plane,
one must specify the sheet (``branch") of the Riemann surface that
one is working in. That is, in the effective action/potential approach
at least, {\it a priori} impossible: there is no reason why one
branch should be preferred to another. Moreover, equations,
similar to Eq. (\ref{e:okop}), in models with fermions \cite{vd99}
have been found to contain roots on the real axis of the second-,
as well as of infinitely many other lower Riemann sheets. In the
present case there are bound to be even more roots
as there are two thresholds and two sets of infinitely many
Riemann sheets. This (``sheet") ambiguity in the effective potential
formalism can be resolved by referring to its connection to the
BS equation. So we turn to the study of analytic properties of the
Gaussian BS equation.

\section{Analyticity of the Gaussian BS equation}
\label{sec:n/d}

We shall show the exact equivalence of the solutions to the Gaussian
BS equation and the so-called N/D equations in S-matrix theory.
N/D equations are one way of implementing the (two-body) unitarity
and causality conditions in a relativistic setting, which, in turn
translate into analytic properties of the scattering amplitude. These
analytic properties are important as they tell us what branch of
the equation to solve and the solution's physical interpretation
(bound state, resonance, ``anti-bound state")
\cite{gas66,nishi74,frau63,collins68}.

The broken-symmetry connected four-point Green function, Eq. (3.12)
in Ref. \cite{issei01}, for the scattering of two non-identical
(``pion-sigma" scattering) scalar particles has the same generic
form of a geometric series as in the symmetric phase, see
Ref. \cite{bg80}, but with an additional pole term in the
``potential" due to the ``elementary'' (massive) ``pion"
excitation, see Eq. (3.11) and Fig. 4 in Ref. \cite{issei01}.
Such a pole term is known in the S-matrix literature as the
Castillejo-Dalitz-Dyson (CDD) pole.
It has been known for some time \cite{zac66} that such a geometric
progression of Feynman diagrams corresponds to the
solution of the (S-wave) N/D equations in the $s$-channel
\begin{eqnarray}
D_{\pi}(s) = {N_{\pi}(s) \over{1 + {1 \over \pi}
\int {d~t \over{t - s - i \epsilon}} N_{\pi}(t)
{\rm Im} \Pi_{\pi}(t)}}
\label{e:N/D}
\end{eqnarray}
The solutions to the N/D equations are not unique, however, the
arbitrariness showing up in the form of so called
Castillejo-Dalitz-Dyson (CDD) poles. The position of a CDD pole,
and the coefficient multiplying it are arbitrary in the usual
S-matrix, or ``bootstrap" approach, but in our approach they are
completely determined by the Gaussian approximation to the
underlying $\sigma$ model Lagrangian.
The physical interpretation of CDD poles used to be controversial,
but the present-day consensus is that they
correspond to elementary particles/fields
in the theory, which conjecture is confirmed
by our results in the Gaussian approximation.

We may rewrite the pion-channel kernel (``polarization function")
$\Pi_{\pi}(s)$ of the Bethe-Salpeter equation
\begin{eqnarray}
\Pi_{\pi}(s) &=& I_{M\mu}(s) \ ~~,
\label{e:polb}
\end{eqnarray}
in the ``dispersive'' form (see Eqs. (\ref{e:eye1})
\begin{eqnarray}
I_{M\mu}(s)
&=& i \int {d^{4} k \over (2 \pi)^{4}}
{1 \over {\left[k^{2} - M^{2} + i \epsilon \right]
\left[(k - P)^{2} - \mu^{2} + i \epsilon \right]}}
\nonumber \\
&=& I_{M\mu}(0) -
{s \over{(4 \pi)^2}} K_{M\mu}(s)
= {1\over{2 \lambda_{0}}}
\left({\mu^{2} \over{\mu^{2} - M^{2}}}\right)
 - {s \over{(4 \pi)^2}} K_{M\mu}(s)
\nonumber \\
&=&
{1\over{2 \lambda_{0}}}
\left({\mu^{2} \over{\mu^{2} - M^{2}}}\right)
- {s \over{16 \pi^3}}
\int {d~t \over{t - s - i \epsilon}}
{\rm Im}~K_{M\mu}(t) ~,\
\label{e:eye2}
\end{eqnarray}
where $s = P^2$ and the real and imaginary parts are
\begin{eqnarray}
{\rm Im}~K_{M\mu}(s)
&=&
{1 \over s} {\rm Im}~I_{M\mu}(s)
\nonumber \\
&=& {\pi \over s}
\sqrt{\left(1 - {(M - \mu)^2 \over s} \right)
\left(1 - {(M + \mu)^2 \over s} \right)}
\theta(s - (M + \mu)^2)
\nonumber \\
{\rm Re}~K_{M\mu}(s)
&=&
{2 \over s} \Bigg[
\left({M^2 - \mu^2 \over 2s}\right) \log{M \over \mu} +
{1 \over 2} \left( 1 +
\left({M^2 + \mu^2 \over M^2 - \mu^2} \right) \log{M \over \mu}
\right)
\nonumber \\
&-&
\sqrt{\left(1 - {(M - \mu)^2 \over s} \right)
\left(1 - {(M + \mu)^2 \over s} \right)}
{\rm tanh}^{-1}
\sqrt{s - {(M + \mu)^2} \over{s - {(M - \mu)^2} }}
\Bigg]~. \
\label{e:eye1}
\end{eqnarray}
The (momentum) $s$ dependent part of this integral is an analytic
function in the cut complex $s$ plane. There are in general two
logarithmic branch cuts (one stretching from $M + \mu$ to $+
\infty$, another from $M - \mu$ to $- \infty$, though on the first
(``physical") sheet only the right-hand-side cut appears)
determining a Riemann surface with infinitely many sheets. This is
rather different from the corresponding nonrelativistic case which
has only one (square root) cut with two sheets.

Comparing Eq. (3.12) in Ref. \cite{issei01} with Eq. (\ref{e:N/D})
above, it becomes clear that the form of the Gaussian approximation
$\pi$ propagator demands that the numerator function $N_{\pi}(s)$
equal the pion channel ``potential" $V_{\pi}(s)$, Eq. (3.11) in
Ref. \cite{issei01} and the form of the denominator in  Eq.
(\ref{e:N/D}):
\begin{eqnarray}
N_{\pi}(s) &=& V_{\pi}(s) =  2 \lambda_{0}
\left(1 + {M^{2} \over s - \mu^{2}} \right)
\label{e:N1}
\\
{1 \over \pi} \int {d t~ N_{\pi}(t) \over{t - s - i \epsilon}}
{\rm Im} I_{M\mu}(t) &=& N_{\pi}(s) \left[I_{M\mu}(0) + {s \over
\pi} \int {d t \over{t(t - s - i \epsilon)}} {\rm Im}~I_{M\mu}(t)
\right] . \label{e:N2}
\end{eqnarray}
Equation (\ref{e:N1}) tells us that $s = \mu^2$ is the position,
and $2 \lambda_{0} M^{2}$ is the coupling strength of the CDD pole,
whereas Eq. (\ref{e:N2}) dictates the value of the ``subtraction
constant"
\begin{eqnarray}
\Pi_{\pi}(0) &=& I_{M\mu}(0)
= {I_{0}(M) - I_{0}(\mu)\over{M^{2} - \mu^{2}}}
\nonumber \\
&=&
{1 \over{2 \lambda _{0}}}\left(1 -
{M^{2} - \varepsilon/v \over{M^{2} - \mu^{2}}} \right)\ ~~,
\label{e:sub}
\end{eqnarray}
that is, in turn, fixed by the gap equations.

Since the N/D approximation is unitary by construction we conclude
that the $s$ channel Gaussian BS scattering amplitude is also unitary.
By the same token, one can show that the $\sigma$ channel propagator
can be written as a solution to the matrix N/D equations, see
Ref. \cite{nishi74}. The subtraction constants are fixed as in
Ref. \cite{issei01}.

Now that we have established analytic properties of the Gaussian
BS equation, we may look for its solutions. As noted above,
physically interesting solutions are to be found as follows: (1)
bound states on the real axis of the physical sheet, below all
thresholds; (2) resonances in the fourth quadrant of the ``second"
sheet; (3) ``antibound states" on the real axis of the second
sheet, below both thresholds. There are also new kinds of S-matrix
singularities in the relativistic quantum field theory (QFT) that
do not appear in nonrelativistic quantum mechanics. One such possible
new relativistic singularity is the CDD ``pole'' (or ``zero'' in the
denominator). CDD poles are associated with ``elementary''
particles/fields in the theory, see  p. 400 in Ref. \cite{gas66}.
In our case this means a field in the $\sigma$ model Lagrangian,
If we remember that the linear $\sigma$ model has been shown to be
the low-energy limit of the Nambu--Jona-Lasinio (NJL) chiral quark
model, we may say that the ``elementary'' particles here are just
the bare NJL quark-antiquark states. These states lead to
real poles only in the weak meson-meson coupling limit, as we shall
show below. Thus the $\sigma$ resonance may become a stable state
only in the limit of weak interactions.

It should be noted, however, that $I_{MM}(s), I_{\mu \mu}(s)$ are
analytic functions with imaginary parts above the corresponding
thresholds, so that Eq. (\ref{e:mass3}) actually implies two
equations: one for the real and one for the imaginary part.
Usually only the real part is considered, however. That is all
right if the (real part of the) root lies below all thresholds. If
it does not, as in our case, one must look at the equation for the
imaginary part, as well. As one moves away from the real $s$ axis,
each equation yields a line of roots in the complex $s$ plane. The
intersection of the two lines (the real and imaginary roots) then
yields the position of the pole.
Only the pole on the second lower sheet
\footnote{$I_{MM}(s), I_{\mu \mu}(s)$ have
logarithmic branch points and therefore infinitely many sheets, in
contrast with the nonrelativistic case where the branch points are
of the square root type, with only two sheets.}, if it exists at
all, determines the mass and width of the resonance.

We have numerically solved the real part of the scalar meson
mass/Gaussian BS equation on the real axis of the physical sheet,
looking for bound states. The results are shown in Fig. 9 of Ref.
\cite{issei01} . Note the double-valuedness of the solutions. The
question arises: can one ascribe particle-like properties to the
heavy branch of the solution. Short of a pole search in the second
sheet, that question can only be answered by calculating the K\"
all\' en-Lehmann spectral function.

\section{K\" all\' en-Lehmann Spectral function}
\label{sec:kallen}

As argued in Sect. III, the Gaussian pion propagator $D_{\pi}(s)$
is an analytic function in the cut $s$-plane and, as such, it allows a
dispersive, or K\" all\' en-Lehmann representation
\begin{eqnarray}
D_{\pi}(s) = - \int dt {\rho_{\pi}(t) \over{t - s - i \epsilon}}~,
\label{e:disp1}
\end{eqnarray}
where $$\rho_{\pi}(s) = - {1 \over \pi} {\rm Im} D_{\pi}(s)~,$$ is
the spectral density function. The latter represents the mass
distribution of physical excitations in this channel. In Ref.
\cite{vd98} we have explicitly shown that in the pion ($J^{P} = 0^{-}$)
channel and in the chiral limit, the spectral function
\begin{eqnarray}
\rho_{\pi}(s) &=&
a \delta(s) + c(s) \theta(s - (M + \mu)^2)~,
\label{e:disp2}
\end{eqnarray}
contains only one Dirac delta function, instead of
two, as naively expected.
In Fig. 1 we show the same spectral density in the nonchiral
case (explicitly broken O(4) symmetry) and again find only one Dirac
delta function, this time at $\sqrt{s} = 140$ MeV.
Another way of saying this is that the
strength with which the state at $s = \mu^{2}$ appears in the
spectrum is zero, i.e., the state decouples from the
single-particle spectrum. The heavier excitations $\phi_{i}, i = 1,2,3$
correspond to unstable quasi-particles \cite{ab97} that decay
into an odd number of {\it lighter} Goldstone bosons. Thus there is
no particle doubling in the Gaussian approximation, contrary to
suggestions e.g. by T\" ornqvist \cite{torn98}.

Similar comments hold for the $\sigma$ sector, the analysis being
more complicated due to two different kinds of
intermediate state being possible there. The sigma channel is
phenomenologically more interesting than the pion one because that
is where many experimental ``supernumerary'' states have been
observed. In Figs. 2, 3, and 4 we show the scalar K\" all\' en-Lehmann
spectral density
$$\rho_{\sigma}(s) = - {1 \over \pi} {\rm Im}(D_{\sigma}(s))~,$$
at various free parameter values. Here the scalar propagator is
defined as
\begin{eqnarray}
D_{\sigma}(s) &=&
{- 2 \lambda_{0} \over{(s - M^2){\cal D}(s)}},
\label{e:sigma}
\end{eqnarray}
the remaining terms in the $D_{ij}$ matrix elements contribute
to the effective $\sigma \phi_{i} \phi_{i},~ i = 0, 1, 2, 3$
vertex form factors. Once again we find only one
bump, or Dirac delta function in the density of states, depending on
the coupling strength $\lambda_{0}$ and other free parameters.

We showed in Fig. 8 of Ref. \cite{issei01} that the scalar channel
(dressed $\sigma$ meson) pole
position is always shifted downward from the ``elementary"  sigma
field's ($\phi^{0}$) CDD pole at $s = M^2$, in accord with the
variational nature of the Gaussian approximation. As the coupling
constant $\lambda_{0}$ drops below some critical value
$\lambda_{c}$, which is a function of the masses $\mu, M$ and the
cutoff $\Lambda$, the $\sigma$ meson mass becomes purely
real as its position drops below the $4 \mu^2$ threshold, see
Fig. 2. Thus, within the Gaussian approximation, the $\sigma$ meson
can be (quasi-)stable at weak couplings, but at moderate and strong
couplings it is always a broad resonance. The behaviour of the
critical value of the $\sigma$ meson mass $m_{\sigma}^{c}$ as a
function of the cutoff $\Lambda$ can be gleaned from  Fig. 8 in
Ref. \cite{issei01} (and similar graphs for values of $\Lambda$
between 0.4 and 1 GeV): the crossing point of the $m_{\sigma}(M)$
and $2 \mu(M)$ curves lies around $M \simeq$ 300 - 400 MeV
until it disappears altogether for $\Lambda > 0.5$ GeV.

In Ref. \cite{issei01} (see Fig. 9.) we showed that at intermediate
values of the coupling $\lambda_{0}$ and low cutoff $\Lambda$ there
is a second, much heavier solution to the real part of the BS equation
besides the usual light one. Thus there is once again the possibility
of a second pole in the S matrix. Such a conclusion would be premature,
as can be seen in Fig. 3, and 4: there is no enhancement of
the scalar spectral function at high energy, whereas at low $s$
the standard solution pole can be seen as a Dirac delta function
turning first into a narrow then a wider peak as one moves up in energy.
Thus we see that there is no enhancement in the density of states at
the corresponding energy/mass and thus there is no second resonance
peak. We
conclude that the heavy solution exists only in the real part of the
BS equation, whereas the imaginary part does not have a root in the
vicinity.

\section{Summary and Conclusions}

The mean-field method was initially fraught with problems when
applied to the linear $\sigma$ model with spontaneously broken
internal symmetry - the Goldstone theorem did not seem to
``work''. This problem was solved, at the price of opening new
questions \cite{dms96}: The Goldstone boson found in the Gaussian
approximation \cite{dms96} turned out to be a composite state that
apparently co-exists with the massive ``elementary" state with
identical quantum numbers. A similar situation occurs in the
scalar sector. At first this looks like a ``doubling'' of the
number of scalar states, sometimes invoked in the phenomenological
literature on the $\sigma$ meson \cite{torn98}.
We have showed here that
this doubling is only apparent: we investigated the question of
the particle content of the $\phi^4$ theory in the Gaussian
approximation by employing the K\" all\' en-Lehmann representation
\cite{hat92}. Thus we found that the massive ``elementary" pion
does not appear in the K\" all\' en-Lehmann spectral function as an
excitation in the pseudoscalar channel. This is the same as saying
that the
Castillejo-Dalitz-Dyson (CDD) ``pole'' in the $\pi$ channel has a
vanishing residue in the MFA to the linear sigma model. Similarly,
only one state has been found to exist in the $\sigma$ channel of
the MFA. Moreover, we showed that the mass of the composite $\sigma$
state agrees exactly with that calculated in the CJT formalism.
Another interpretation, (that might be only semantically different
from the above one) was derived by using operator many-body
(``quasi-particle RPA'') methods \cite{ab97,fw71}. In that
theoretical framework the ``lighter'' states are particles and the
``heavier'' ones are so-called quasi-particles that appear due to
the interactions of the particles.

In summary, we have: 1) established equivalence of the standard
Gaussian approximation with the effective potential method; 2)
shown equivalence to the N/D equations and thus proven unitarity.
3) solved the resulting equations and investigated the theories
particle content via the K\" all\' en-Lehmann representation.

One of the authors [V.D.] wishes to thank Prof. Paul Stevenson for
valuable conversations and correspondence relating to the Gaussian
approximation in general and some of the problems treated here in
particular. The same author would also like to acknowledge a COE
Professorship at RCNP for the year 2000/1 at which time most of
this work was done.

\newpage
\vspace{1cm}

\begin{figure}
\caption{The isovector pseudoscalar (pion channel) spectral
density $\rho_{\pi}(s)$ in the Gaussian approximation to the O(4)
linear sigma model as a function of the CM energy $\sqrt{s}$ for
various values of the free parameters. The vertical line at
$\sqrt{s}$ = 140 MeV represents the Dirac delta function. Note the
absence of other delta functions. The threshold of the continuum
is at $\sqrt{s} =  M + \mu$.} \label{f:1}
\end{figure}

\vspace{1cm}

\begin{figure}
\caption{The isoscalar scalar (sigma channel) spectral density
$\rho_{\sigma}(s)$ in the Gaussian approximation to the O(4)
linear sigma model as a function of the CM energy $\sqrt{s}$ for
$\Lambda$ = 1 GeV and various values of the variational parameter
(mass) $M$, i.e. of the coupling constant $\lambda_{0}$.
The vertical line
represents the Dirac delta function. The first threshold of the
continuum is at $\sqrt{s} = 2 \mu$, the second threshold
($2``\sigma"$) is at $\sqrt{s} =  2 M$ where a cusp in the spectral
density can be seen. In both cases the physical $\sigma$ mass lies
below the lower threshold at $\sqrt{s} = 2 \mu$, so the $\sigma$ 
meson is stable.} \label{f:2}
\end{figure}

\vspace{1cm}

\begin{figure}
\caption{As in Fig. 2, but for cutoff $\Lambda$= 0.4 GeV and
three values of the variational parameter $M$ = 330, 370 and
500 MeV (with $\mu$= 146, 156, and 165 MeV,  respectively).
In one case the physical $\sigma$ mass lies
below the lower threshold at $\sqrt{s} = 2 \mu$, in the other 
two it lies above their respective thresholds, so accordingly 
the $\sigma$ meson is either stable or a resonance.
Note the widening of the $\sigma$ resonance peak as its mass
increases, and the coupling constant grows $\lambda_0$ with it.}
\label{f:3}
\end{figure}

\vspace{1cm}

\begin{figure}
\caption{Same as in Fig. 3, but rescaled to show higher
values of $\sqrt{s}$:
note the absence of any enhancement in the region of
the ``second solution" (the cusp is due to the second
threshold).}
\label{f:4}
\end{figure}


\begin{thebibliography}{10}

\bibitem{bg80}
T. Barnes and G.I. Ghandour, Phys. Rev. D {\bf 22}, 924 (1980);
see also eds. L. Polley and D.E.L. Pottinger,
{\em Variational Calculations in Quantum Field Theory}, World
Scientific, Singapore (1987).
\bibitem{hat92} R.J. Rivers
{\em Path Integral Methods in Quantum Field Theory},
Cambridge University Press, Cambridge (1987), and
B. Hatfield,
{\em Quantum Field Theory of Point Particles and Strings},
Addison-Wesley, Reading (1992).
\bibitem{issei01}
Issei Nakamura and V. Dmitra\v sinovi\' c,
\newblock Prog. Theor. Phys. {\bf 106}, 1195 (2001).
\bibitem{amcam96} G. Amelino-Camelia, J. D. Bjorken, and
S. Larsson, Phys. Rev. D {\bf 56} 6942 (1997);
 S. Chiku and T. Hatsuda, Phys. Rev. D {\bf 57},
R 6 (1998), and Phys. Rev. D {\bf 58}, 076001, (1998);
H.-S. Roh, and T. Matsui, Eur. Phys. J. A {\bf 1}, 205 (1998);
J. T. Lenaghan, D. H. Rischke, J. Schaffner-Bielich, Phys.Rev.{\bf D 62},
085008 (2000);
Y. Nemoto, K. Naito and M. Oka, Eur. Phys. J. A {\bf 9}, 245 (2000).
\bibitem{sat87}
P. M. Stevenson, B. All\` es and R. Tarrach,  Phys. Rev. {\bf D
35}, 2407 (1987).
\bibitem{okop96} A. Okopi\~ nska, Phys. Lett. {\bf B 375}, 213 (1996).
\bibitem{gas66} S. Gaziorowicz, {\it Elementary Particle Physics},
      (J. Wiley and Sons, New York, 1966).
\bibitem{nishi74} K. Nishijima,
{\it Fields and Particles}, (W.A. Benjamin, Reading, MA, 1969).
\bibitem{frau63} S. C. Frautschi, {\it Regge Poles and S-matrix Theory},
      (W.A. Benjamin, New York, 1963).
\bibitem{collins68} P. D. B. Collins and E. J. Squires,
{\it Regge Poles in Particle Physics},
(Springer, New York, 1968).
\bibitem{cjt74} J.M. Cornwall, R. Jackiw and E. Tomboulis, Phys. Rev. D
{\bf 10} 2428 (1974).
\bibitem{iz80} C. Itzykson and J.B. Zuber, {\it Quantum Field Theory},
(McGraw-Hill, New York 1980).
\bibitem{fw71} A. Fetter and J.D. Walecka,
{\it Quantum Theory of Many-Body Systems}, (McGraw-Hill, New York 1971).
\bibitem{dms96} V. Dmitra\v sinovi\' c, J. A. McNeil and J.
Shepard, Z. Phys. {\bf C 69}, 359 (1996).
\bibitem{vd98} V. Dmitra\v sinovi\' c, Phys. Lett. {\bf B 433}, 362 (1998).
\bibitem{vd99} V. Dmitra\v sinovi\' c, Phys. Lett. {\bf B 451}, 170 (1999).
\bibitem{zac66} F. Zachariasen, p. 97 - 99 in
{\it Recent Developments in Particle Physics}, editor M. J. Moravcsik,
      (Gordon and Breach, New York, 1966).
\bibitem{ab97} H. W. L. Aouissat, O. Bohr, and J. Wambach,
Mod. Phys. Lett. {\bf A 13}, 1827 (1998).
\bibitem{torn98} N. T\" ornqvist,
Phys. Lett. {\bf B 426}, 105 (1998).
\end{thebibliography}
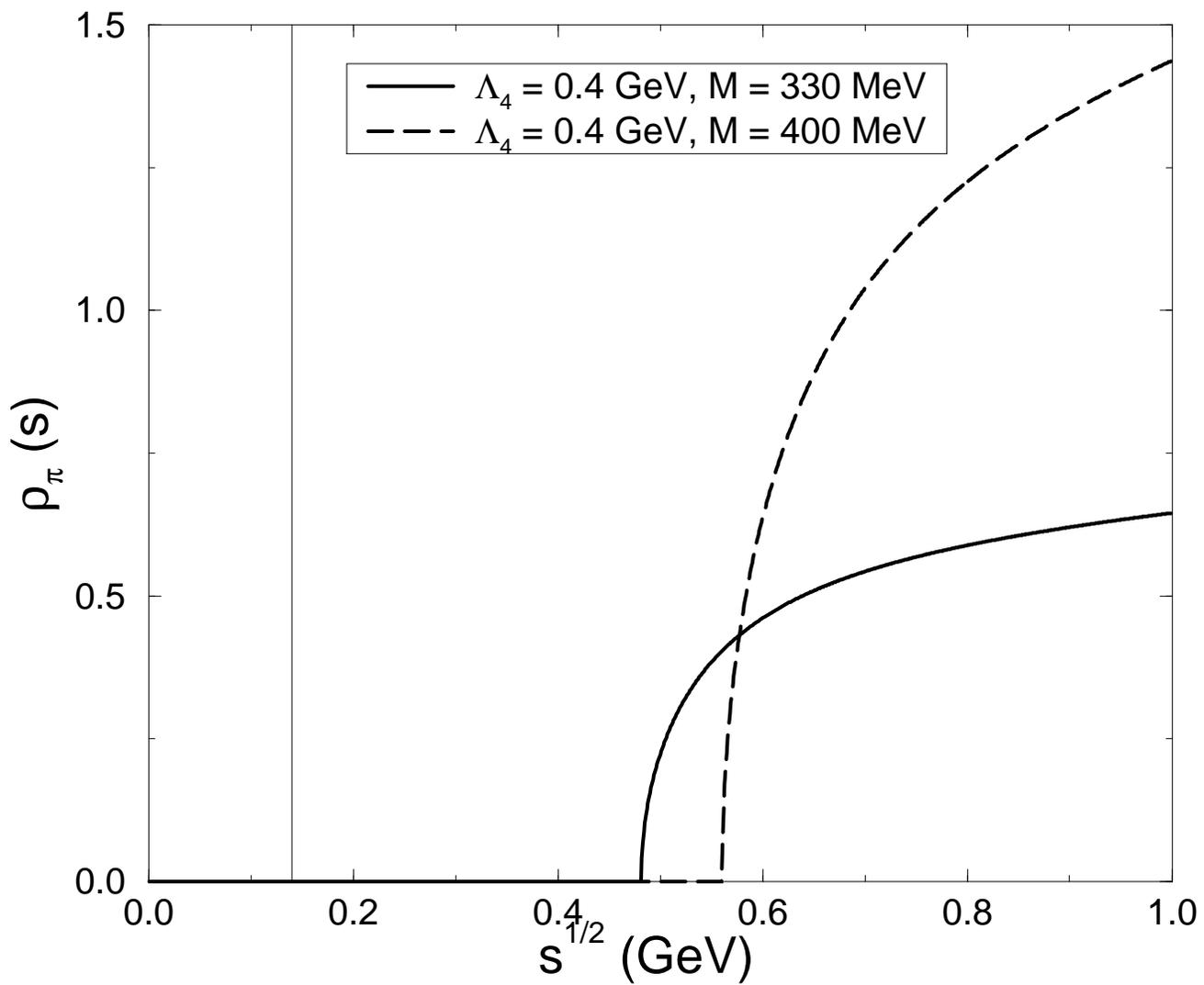
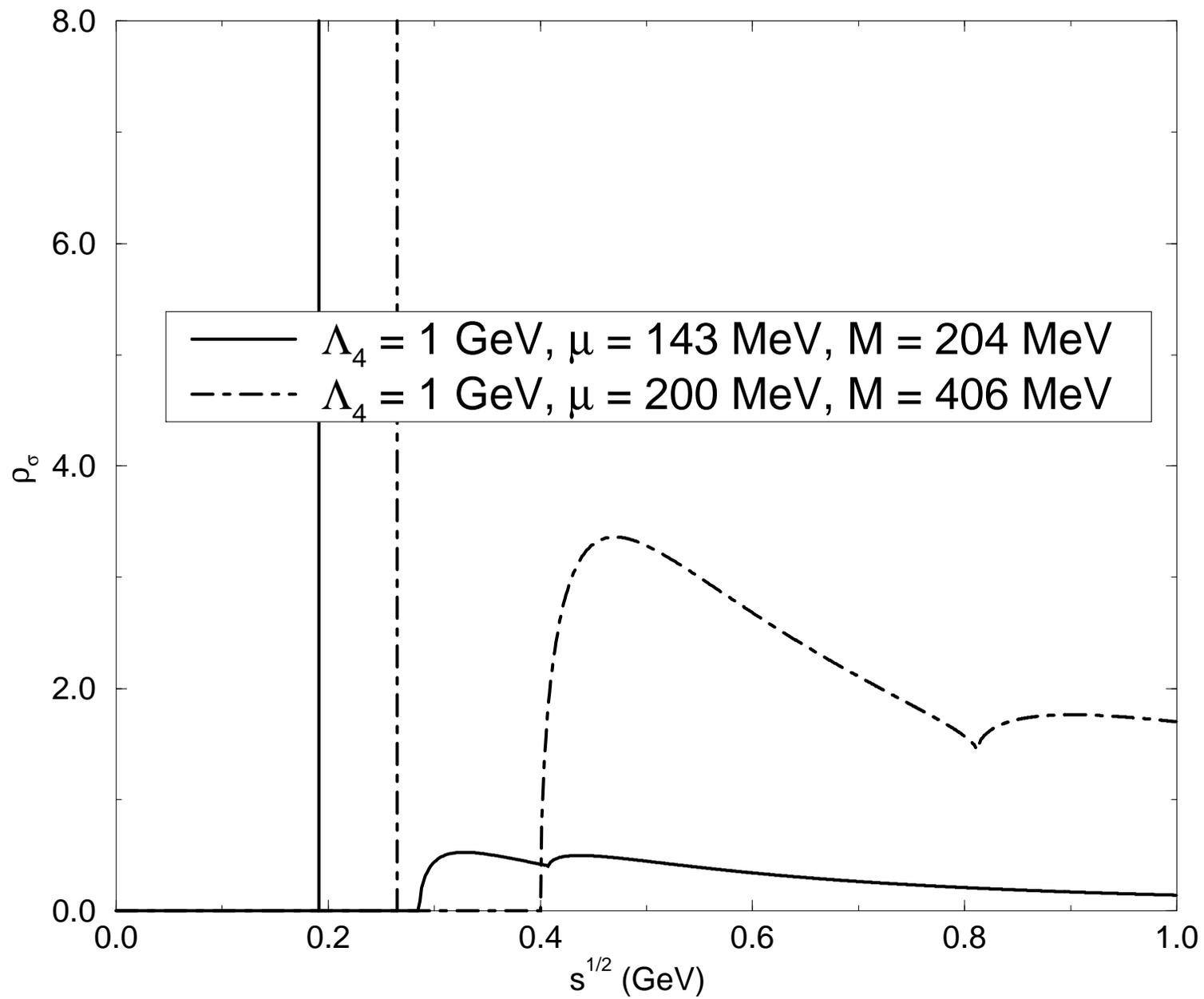
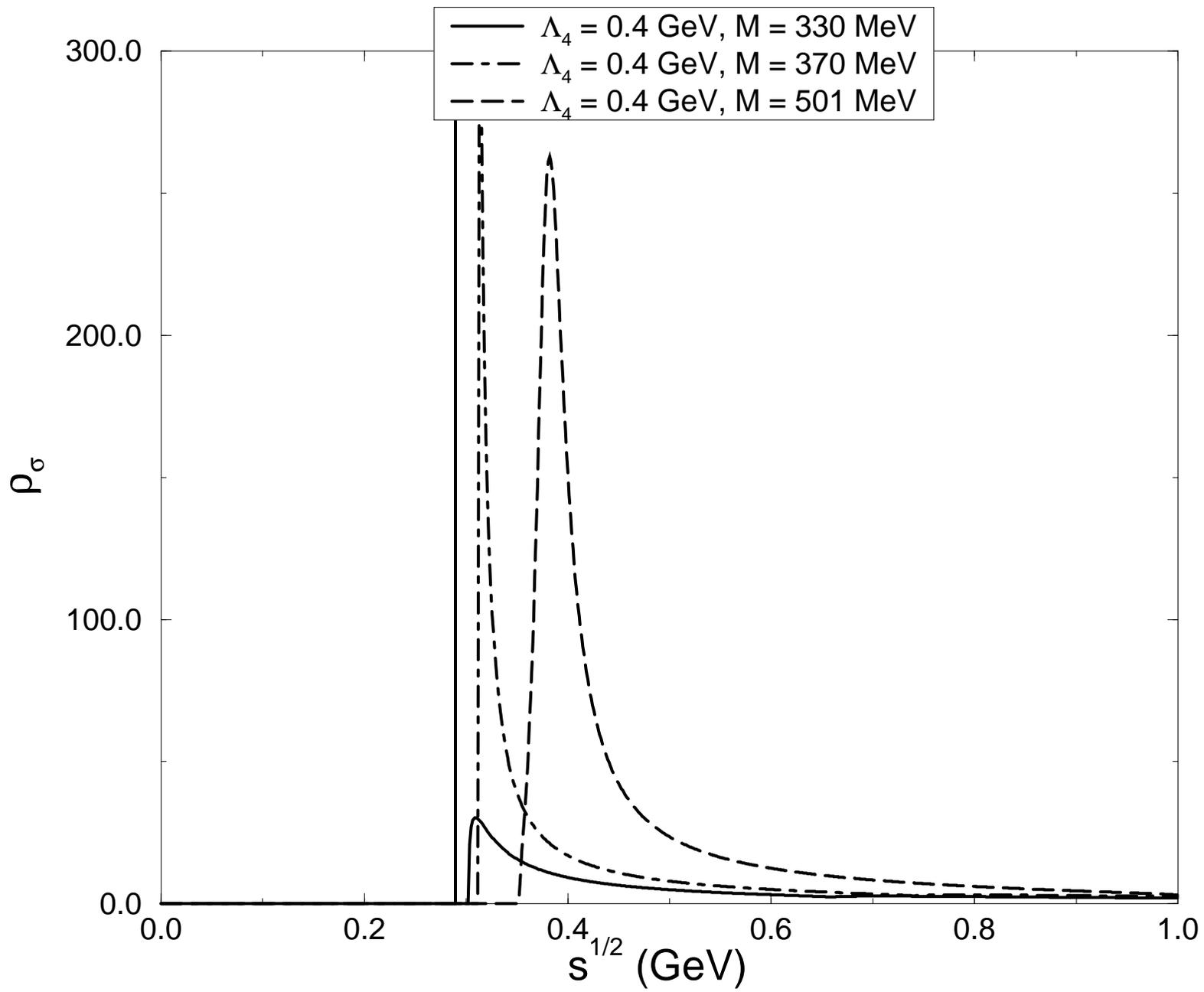
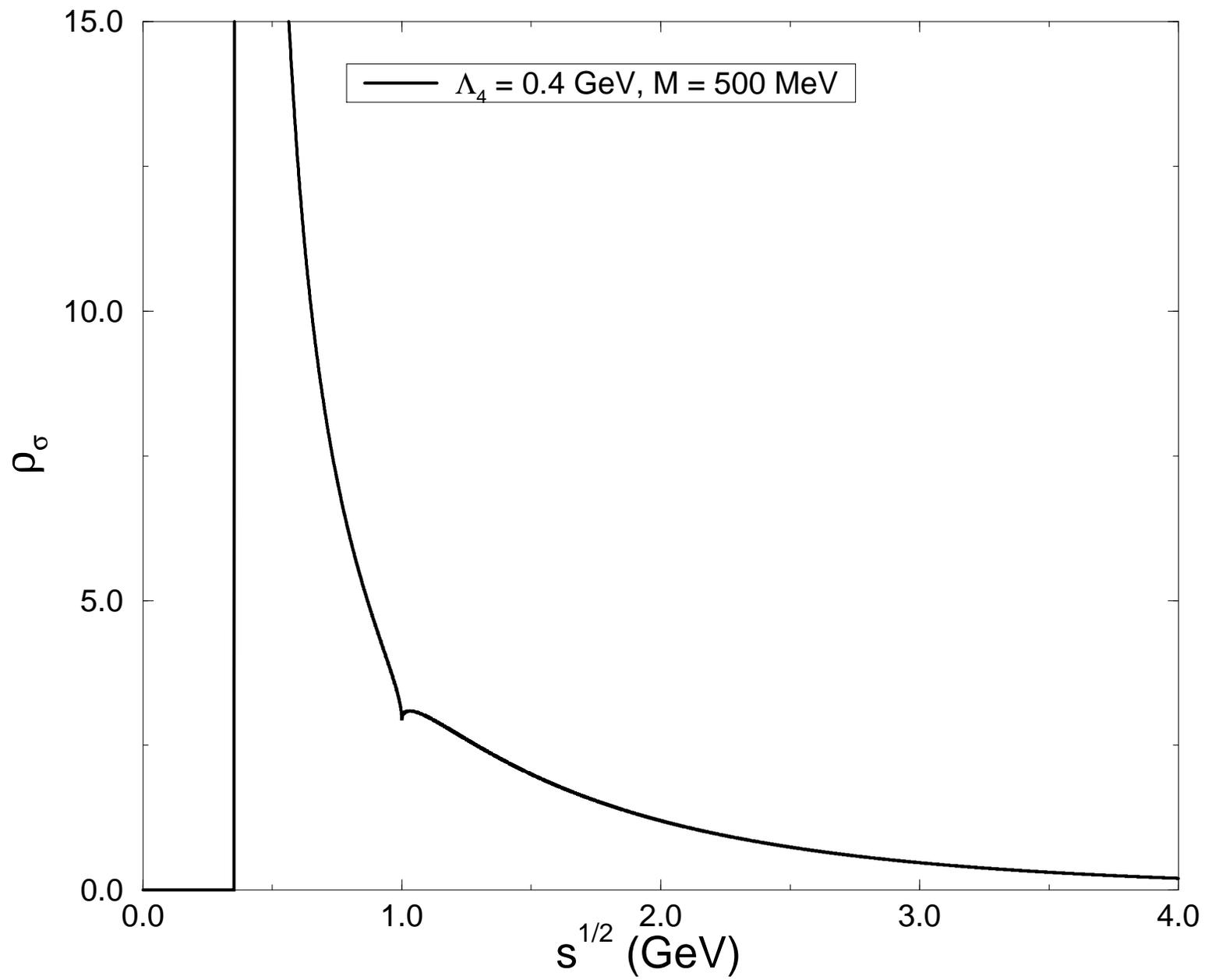
\end{document}